\documentclass[apj]{emulateapj}
\usepackage{amsmath}
\newcommand{\Dopamp}{1.86 $\pm$ 0.25 m s$^{-1}$}
\newcommand{\mass}{1.87$^{+0.27}_{-0.26}$ M$_{\oplus}$}
\newcommand{\density}{6.0$^{+1.9}_{-1.4}$ g cm$^{-3}$}
\newcommand{\ironfrac}{$32 \pm 26$\%}

\begin{document}

\title{Determining the Mass of Kepler-78\lowercase{b} with Nonparametric Gaussian Process Estimation}
\author{Samuel K.\ Grunblatt}
\author{Andrew W.\ Howard}
\affil{Institute for Astronomy, University of Hawaii,
2680 Woodlawn Drive, Honolulu, HI 96822}
\author{Rapha{\"e}lle D.\ Haywood}
\affil{SUPA School of Physics and Astronomy, University of St Andrews, 
North Haugh, St Andrews, Fife KY16 9SS, United Kingdom}

\email{Email: skg@ifa.hawaii.edu}
\shorttitle{Kepler-78b Mass Measurement with Gaussian Process}
\shortauthors{Grunblatt et al.}

\begin{abstract}
Kepler-78b is a transiting planet that is 1.2 times the radius of Earth and orbits a young, active K dwarf every 8 hours. The mass of Kepler-78b has been independently reported by two teams based on radial velocity measurements using the HIRES and HARPS-N spectrographs. Due to the active nature of the host star, a stellar activity model is required to distinguish and isolate the planetary signal in radial velocity data. Whereas previous studies tested parametric stellar activity models, we modeled this system using nonparametric Gaussian process (GP) regression. We produced a GP regression of relevant \textit{Kepler} photometry. We then use the posterior parameter distribution for our photometric fit as a prior for our simultaneous GP + Keplerian orbit models of the radial velocity datasets. We tested three simple kernel functions for our GP regressions. Based on a Bayesian likelihood analysis, we selected a quasi-periodic kernel model with GP hyperparameters coupled between the two RV datasets, giving a Doppler amplitude of \Dopamp and supporting our belief that the correlated noise we are modeling is astrophysical. The corresponding mass of \mass\ is consistent with that measured in previous studies, and more robust due to our nonparametric signal estimation. Based on our mass and the radius measurement from transit photometry, Kepler-78b has a bulk density of \density.  We estimate that Kepler-78b is \ironfrac\ iron using a two-component rock-iron model.  This is consistent with an Earth-like composition, with uncertainty spanning Moon-like to Mercury-like compositions. 


\end{abstract}

\section{Introduction}

Measuring the radial velocity (RV) of a planet's host star is the most common method to measure the mass of a planet. The planetary RV signal, or Doppler amplitude, is directly related to the planet mass. If we know the mass of the star, we can find the mass of the planet from its Doppler amplitude. If the planet radius is also known, we can then calculate the planet's density and estimate a composition. Precise density and composition information is available only for a handful of transiting rocky planets. 

However, stellar activity can produce spurious RV signals larger than some planetary signals. In young, spotted stars, this activity can cause RV variations much larger than most observed planetary RV signals \citep{hillenbrand2015}, and even in relatively quiet stars such as the Sun, these variations are at least one order of magnitude larger than the RV signal of Earth \citep{meunier2010}. Before we can have any hope of confirming the discovery of an Earthlike planet around any star, we must account for the contribution of stellar activity.

Discovered in 2013, the radius of Kepler-78b was measured to be 1.16 $\pm$ 0.16 R$_{\oplus}$ \citep{sanchis-ojeda2013}. Subsequently, the mass of the planet was measured and reported simultaneously by two teams. Based on the radius of the planet and the age of the star established by \citet{sanchis-ojeda2013}, the expected Doppler amplitude was 1-2 m s$^{-1}$, while the expected spot-induced RV signal was approximately an order of magnitude larger ($\approx$10 m s$^{-1}$). \citet[][H13]{howard2013} and \citet[][P13]{pepe2013} independently observed and modeled the RV signal of Kepler-78 parametrically to remove any contribution from correlated noise activity while measuring the planetary Doppler signal. H13 used a sum of sinusoids at the stellar rotation period and its harmonics to model the correlated noise activity, accounting for rotational variability while neglecting any component of the predominant signal that was not periodic at precisely the rotation period of the star or its aliases. 

P13 noted the fact that the planetary and stellar RV signal timescales differ by over an order of magnitude: the orbital period of the planet is only 8.5 hours, whereas the stellar activity-induced signals are modulated by the stellar rotation period of 12.8 days. This allowed P13 to test a floating chunk offset model in which the free parameters were the planetary Doppler amplitude and an RV zero point to represent the noise signal, assumed constant for each night of observations. However, P13 found that the evidence ratio for their floating chunk offset model was significantly lower than the sum of sinusoids model, given the large number of free parameters, and used the parametric, sinusoidal model for their reported mass measurement. \citet{howard2013}used Keck/HIRES and measured the mass of Kepler 78b to be 1.69  $\pm$ 0.41M$_\oplus$, while \citet{pepe2013} used TNG/HARPS-N and measured a mass of 1.86$^{+0.38}_{-0.25}$ M$_\oplus$. Despite the difference in their methods and observational techniques, the H13 and P13 planet mass measurements are consistent, attesting to the robustness of the results.

Further exploration into nonparametric estimation is justified by the limited scope of the purely parametric models tested previously. We draw attention to the comparable case of HST transmission spectroscopy of hot Jupiter HD 189733, studied by \citet{swain2008}, \citet{gibson2011} and \citet{gibson2012}. \citet{gibson2012} note that both previous studies of the transmission spectroscopy used linear basis functions to account for systematic errors, and argue that this is not sufficient to account for instrumental systematics, and therefore provides an unrealistic treatment of the uncertainties. \citet{gibson2012} reanalyzed the spectroscopic data with a Gaussian process to marginalize ignorance of the functional form of the systematics, giving larger but more robust errors. Similarly, we have reanalyzed the RV data of Kepler-78b with several Gaussian process estimators to find a model of the RVs with the highest evidence, resulting in more robust uncertainties of the mass, density, and composition of Kepler-78b. In this case, we are using Gaussian process to describe a signal of astrophysical origin, rather than instrumental, which we confirm through tests described in our Results section.\


In this study, we combine the RV measurements from H13 and P13. We describe each radial velocity dataset with a Gaussian process (GP) regression combined with a Keplerian orbit signal at the orbital period and phase of the planet (both known precisely from the transit photometry). GP regression is a nonparametric method for modeling correlated RV noise, and can provide a more robust noise model because of its flexibility over parametric models. The observations analyzed are described in $\S$2. Our analysis of the combined RV dataset is described in depth in $\S$3, starting with a review of Gaussian process regression and the benefit of a nonparametric model in $\S$ 3.1-3.3, and a focus on how different GP models were tested and chosen for the data in sections $\S$ 3.4 and 3.5. In $\S$4, we report our results for the planet Doppler amplitude and mass, and discuss the selection of the chosen analysis model. Using the known properties of the host star (summarized in Table \ref{tbl-1}) and the planet radius, we recalculate the planetary density. In $\S$5, we compare our mass measurement and density calculation to previous results and discuss possible composition scenarios for Kepler-78b. We explore the applications of this technique to other RV datasets in $\S$6.


\begin{deluxetable}{ccrrrrrrrrcrl}[h]
\tabletypesize{\scriptsize}
\tablecaption{Stellar Properties\label{tbl-1}}
\tablewidth{0pt}
\tablehead{
\colhead{Property} & \colhead{Value}
}
\startdata
Name & Kepler-78/KIC 8435766/Tycho 3147-188-1 \\
Age & 625 $\pm$ 150 million years\\
 V$sin(i)$ &  2.6 $\pm$ 0.5 km s$^{-1}$\\
Mass, $M_{\mathrm{star}}$ & 0.83 $\pm$ 0.05 $M_{\odot}$\\
Inclination, $i$ & $75.2^{+2.6}_{-2.1}$ degrees\\
Rotation period $P_{\mathrm{rot}}$ & $\approx$ 12.5 days\\ 
\enddata 
\tablecomments{The values in Table \ref{tbl-1} are have been taken from \citet{howard2013}.}

\end{deluxetable}

\section{Observations}

In order to determine the Doppler amplitude of Kepler-78b, it was necessary to first confirm the general structure of the stellar activity by modeling the \textit{Kepler} photometry. These model parameters were then used to provide reasonable initial hyperparameters for the RV analysis. The observations taken are described below. 

\subsection{\textit{Kepler} photometry}

The \textit{Kepler} telescope obtained photometry of approximately 150,000 objects in the \textit{Kepler} field for its four year lifetime from 2009 to 2013. Photometry of Kepler-78 was gathered at a 30-minute cadence. In this study, we train a GP on the photometric light curve of Kepler-78 to determine the evolution and rotation timescales of the stellar activity, hyperparameters of the GP regression kernel. However, as shown in Equation \ref{eq:2}, a matrix inversion is required to calculate the log posterior likelihood of a GP kernel with a given set of hyperparameters, a computationally intensive process with compute time proportional to $N^3$. Therefore, it was necessary to rebin the \textit{Kepler} photometry to one point every 5 hours (averaging every ten points together) in order to find the best-fit hyperparameters of a full quarter of photometry with our computing resources. While this does marginalize over the planetary signal, we are only using the photometry to estimate the underlying components of the stellar activity which vary on timescales much longer than 5 hours, and the photometric effect of the planet is small compared to that of the stellar activity. This photometry, taken from the 16th quarter of \textit{Kepler} observations, is not concurrent with the RV measurements; however, since it was taken only $\approx$55 days before the RV measurements, we can assume that the variation seen in the photometry is at least related to the RV variation, and would exhibit similar structure in the temporal dimension. Figure \ref{fig:bestphotfit} illustrates the binned photometric measurements as well as the GP regression with the best-fit kernel hyperparameters to the photometry.

\subsection{HIRES and HARPS-N RV data}

The RV observations were obtained shortly after the malfunctioning of \textit{Kepler}'s reaction wheels in summer 2013, which prevented the high level of photometric precision achieved in the first era of the spacecraft's lifetime and required that the telescope point at new targets. H13 observed Kepler-78 with HIRES on the Keck I telescope on Mauna Kea. They obtained eight nights of data during each of which approximately 9 or more measurements were obtained: three measurements were taken at successive thirty minute intervals at three separate occasions per night. In addition, five individual measurements were taken on separate nights after the rest of the data were collected to provide a longer baseline over which to evaluate the stellar activity. The dataset consists of 84 RVs over 45 nights. The P13 team observed Kepler-78 with HARPS-N on the Telescopio Nazionale Galilei in the Canary Islands. They were able to obtain 109 measurements over 97 nights. Most of their observations were made in a consecutive six-day period. On subsequent nights only one or two measurements were made per night, giving a total time baseline of approximately three months. Figure \ref{fig:bestGPfit} illustrates the HIRES and HARPS-N measurements with measurement errors shown, with GP regressions overplotted.

\begin{figure*}
\includegraphics[width=\textwidth]{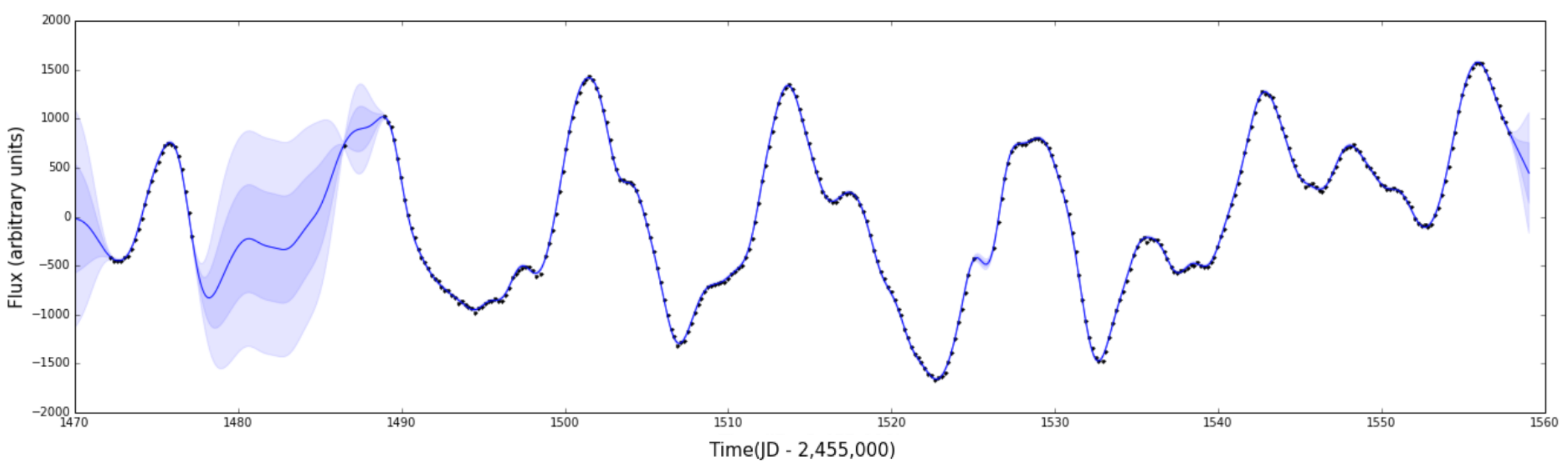}\label{fig:bestphotfit}
\caption{Flux of Kepler-78 versus time, as recorded by the \textit{Kepler} spacecraft during its 16th quarter of observations. The data are plotted as black points, with photometric errors shown. The blue line corresponds to the Gaussian process regression of the photometry with the best-fit kernel hyperparameters, and the shaded regions correspond to 1-$\sigma$ and 2-$\sigma$ uncertainties, as defined by the posterior distributions of the kernel hyperparameters. The \textit{Kepler} data has been shown as fitted with a single quasi-periodic kernel function and a white-noise stellar jitter parameter.}
\label{fig:bestphotfit}
\end{figure*}

\section{Methods}

We model the photometric variations observed by \textit{Kepler} with a GP regression with a quasi-periodic kernel. We then model the radial velocity measurements with a quasi-periodic kernel GP regression, using the kernel period hyperparameter, $\theta_{\mathrm{phot}}$, to train the RV GP period hyperparameter $\theta_{\mathrm{RV}}$. This assumes that the predominant signal observed in both datasets comes from the star, and thus will modulate at the stellar rotation period. When we model the photometric data, we recover a period hyperparameter that agrees with other autocorrelation analyses of the \textit{Kepler} photometry \citep{sanchis-ojeda2013}, supporting this assumption. We create a GP regression of each RV dataset with a quasi-periodic kernel, where the period hyperparameter has been trained on the period hyperparameter of the photometric regression. We then fit all the RV GP hyperparameters, $h$, $\theta$, $w$, and $\lambda$ in the quasi-periodic case, $\sigma$, a white noise parameter, as well as the Doppler amplitude $K$ of a Keplerian signal at the known orbital period and phase of the planet, to the RV data simultaneously. We find that all of our GP + Keplerian models are able to recover the planetary Doppler amplitude $K$ at a value consistent with previous work.

\subsection{Gaussian Process Regression: Concept}

Gaussian process is a nonparametric method to describe a dataset by evaluating correlations between $n$ data points through a covariance kernel. This kernel describes the relationship of each point in the dataset to each other point, and can be expressed as an $n \times n$ matrix (subsequently referred to as the covariance matrix). The kernel is a function of hyperparameters. More complicated kernels can have more hyperparameters that characterize different qualities of the correlations in the data, such as various periods, characteristic amplitudes and length scales, \textit{etc.} 

Gaussian process regression is widely used in the field of machine learning \citep{neal1997, lawrence2003, quinonero2005, wang2008}. \citet{gibson2012} introduced the technique to the field of exoplanets through analysis of transmission spectroscopy to model correlated noise in the instrumental systematics of $HST$/NICMOS, as described in $\S$1. Concurrently with this work, \citet{haywood2014} have demonstrated the technique of GP modeling of RV and photometric signals for the CoRoT-7 planetary system, first modeling the photometry with a GP and then using the photometric GP hyperparameters to train the initial RV GP hyperparameters. \textit{\citet{haywood2014} demonstrated that in the case of CoRoT-7b, parametric spot models such as those used by H13 and P13 gave incorrect masses and uncertainties. Thus, it is important to test many time series techniques, and further explore the novel application of GPs in Doppler analysis.} We expand upon the GP training technique used in \citet{haywood2014} here.

\subsection{Gaussian Process Covariance Kernel Choice}

Finding the best GP regression requires choosing a kernel and initial hyperparameters, evaluating the likelihood of those hyperparameter values, and then iterating through parameter space until the most likely values are found. The squared exponential kernel, for example, defines a covariance matrix through an operator, 

 \begin{equation} \label{eq:1}
\Sigma_{ij} = k(t_{i},t_{j}) = h^2 \mathrm{exp}\bigg[ - \Big(\frac{{t_{i} - t_{j}}}{\lambda}\Big)^2\bigg],
 \end{equation}
where $h$ is the covariance amplitude, and $\lambda$ the covariance length scale. The amplitude observed is described by $h$, while $\lambda$ is a characteristic timescale over which the data is going to be correlated.


We discuss other GP kernels and the inferred physical meaning of their hyperparameters in Table \ref{tbl-3}. 

The logarithm of the posterior likelihood of the GP regression is calculated as 
\begin{multline} \label{eq:2}
\mathrm{log}[\mathcal{L}(\mathbf{r})] = -\frac{1}{2}\mathbf{r}^\mathrm{T}\mathbf{\Sigma}^{-1}\mathbf{r} - \frac{1}{2} \mathrm{log} |\mathbf{\Sigma}| - \frac{n}{2} \mathrm{log}(2\pi),
\end{multline}
where $\mathbf{r}$ is the vector of residuals after removal of the (optional) mean function, $\mathbf{\Sigma}$ is the covariance matrix, and $n$ the number of data points. A prior term, $\mathcal{L}_{\mathrm{prior}}$, can be added to the likelihood to account for any priors placed on the hyperparameters. For example, we apply the Gaussian prior 
\begin{equation} \label{eq:3}
\mathcal{L}_{\mathrm{prior}} = e^{-\frac{1}{2}\Big[\left(\frac{\theta_{\mathrm{true}} - \theta}{\sigma_{\theta}}\right)^2 \Big]},
\end{equation}
to restrict the hyperparameter $\theta$ for the RV data, as the period of the correlated noise signal is equivalent to the stellar rotation period P$_{\mathrm{rot}}$ in both the photometric and RV datasets. We restrict the period hyperparameter $\theta$ in the RV regressions within a Gaussian of width $\sigma_{\theta}$ determined by the posterior distribution of the period hyperparameter found for our photometric GP regression. Prior knowledge of this hyperparameter helps to ensure convergence of the other hyperparameters of the RV regression, as the RV data is not as well sampled as the photometry.

 This likelihood calculation can be used to identify the best fit GP hyperparameters. For more details on how the covariance kernel and parameter boundary conditions were chosen, refer to $\S$ 4.1. For a more complete description of Gaussian process regression and posterior likelihood evaluation, see \citet{rasmussen2006}.

\begin{deluxetable*}{cccc}
\tabletypesize{\scriptsize}
\tablecaption{Gaussian Process Kernel Options \label{tbl-3}}
\tablewidth{0pt}
\tablehead{
\colhead{Name} & \colhead{Mathematical expression} & \colhead{Hyperparameters\tablenotemark{a}} & \colhead{Comments}
}
\startdata
Squared exponential & $h^2 \mathrm{exp}\bigg[ - \Big(\frac{{t_{i} - t_{j}}}{\lambda}\Big)^2\bigg]$ & $h$, $\lambda$ & $h$ amplitude of covariance function,  \\
 & & & $\lambda$ a characteristic timescale \\
Periodic  & $h^2 \mathrm{exp}\bigg[ -\frac{\mathrm{sin}^2[\pi(t_i - t_j)/\theta]}{2w^2} \bigg]$ &  $h$, $\theta$, $w$ & $\theta$ equivalent to P$_{\mathrm{rot}}$, \\ 
 & & & $w$ represents coherence scale, similar to $\lambda$ expressed \\
 & & & as a fraction of $\theta$ dependent on recurrent features\\
Quasi-Periodic &  $h^2 \mathrm{exp}\bigg[ -\frac{\mathrm{sin}^2[\pi(t_i - t_j)/\theta]}{2w^2} - \Big(\frac{{t_{i} - t_{j}}}{\lambda}\Big)^2\bigg]$ & $h$, $\theta$, $w$, $\lambda$ & $w$ coherence scale tied to periodic variation \\
& & & while characteristic timescale $\lambda$ tied to aperiodic variation. \\
\enddata 
\tablecomments{The name of kernel functions and hyperparameters in Table \ref{tbl-3} are taken from \citet{rasmussen2006}.}
\tablenotetext{a}{Each kernel $\mathbf{\Sigma_{ij}}$ can be modified to include an additional hyperparameter, a white noise term $\sigma^2$ by adding one in quadrature: $\mathbf{\Sigma_{ij}} = \mathbf{\Sigma_{ij}}$+ $\sigma^2\textbf{I}_{i}$.}

\end{deluxetable*}

\subsection{Gaussian Process and Stellar Activity}

The presence of magnetic features on the stellar surface induces variation in a star's RV that can mimic a planet's orbital RV signal. Starspots are regions of high magnetic fields on the surface of a star that appear relatively cooler and darker than the surrounding photosphere. As the starspots move across the disk of a star (due to stellar rotation), the flux balance between the redshifted and blueshifted halves of the star is broken, producing a shift in the centroid of the spectral lines. This shift corresponds to an apparent RV signal and change as the spots move across the stellar disk \citep{dumusque2011}. Starspots, as well as networks of strongly magnetized flux tubes known as faculae, inhibit the convective processes taking place on the stellar surface, thus reducing the net blueshift produced by the up flow of hot, bright granules. This has been shown to be the dominant RV effect on the Sun \citep{meunier2010} and on other Sun-like stars \citep{haywood2014}. These activity-induced RV signals are modulated by the stellar rotation and the surface features evolve with time, resulting in a quasi-periodic, RV signal. The lifetimes of surface features are poorly constrained, making this effect difficult to model parametrically. The nonparametric nature of the GP regression provides an ideal framework for studying stellar surface features in RV. We can choose a GP kernel that reflects the frequency structure of the stellar activity. The resultant GP model is flexible enough to account for the evolution of magnetic features on the stellar surface while keeping a statistical ``memory" of the stellar activity patterns.

\subsection{Photometric Model}

We fit the photometric data using GP regressions with three different kernels. We test a squared exponential, periodic and quasi-periodic kernel (Table \ref{tbl-3}). We test kernels with and without an additional white noise term added in quadrature to the likelihood (Eq. 2). The best-fit kernel hyperparameters are found via Markov Chain Monte Carlo analysis powered by the \texttt{emcee} Python package \citep{foremanmackey2013}, described in more detail in the following section. We select the quasi-periodic model with a white noise term based on our assumption that the predominant signal in both datasets is tied to the quasi-periodic variation of the stellar surface, and confirm that the quasi-periodic model is appropriate through a visual check and a reduced $\chi^2$ analysis. We have plotted the data and GP regression with the best-fit kernel hyperparameters in Figure \ref{fig:bestphotfit}. We ensure that these hyperparameters are robust by performing a GP regression to the photometric data after less stringent binning, and recovering the same best-fit kernel hyperparameter values within errors.

 \subsection{RV Model}

We fit our photometric and RV datasets with the same kernel operator. We took the hyperparameter values from the photometric GP regression and used them as initial fit values for the RV GP kernel hyperparameters, with the exception of the hyperparameters $h$ and $\sigma$, as the variation in the photometry and RV related to these quantities is not in equivalent units. We also place a Gaussian prior on $\theta$ based on its posterior distribution determined from the photometric GP regression. We then apply a uniform offset to each set of RV measurements to remove RV zero point differences. 

The likelihood of the GP regression, given certain kernel hyperparameters and a Doppler amplitude value, determines the quality of the model for those given parameters. This likelihood can be calculated as given in Equation 2, where the residuals $\textbf{r}$ can be calculated as
\begin{equation}
\mathbf{r} = \mathbf{v} - K \mathrm{sin}\Big(\frac{2\pi(\mathbf{t} - t_{c})}{P_{\mathrm{orb}}}\Big),
\end{equation}
where $\mathbf{t}$ is the vector of times of all measurements, $\mathbf{v}$ the vector of RV measurements, $t_c$ a time of transit, $P_{\mathrm{orb}}$ the orbital period, and $K$ the planetary Doppler amplitude. \citet{sanchis-ojeda2013} measured $t_c$ to 10$^{-5}$ of a day and and $P_{\mathrm{orb}}$ to 10$^{-7}$ of a day precision, so we can safely treat them as constants for this analysis.

\begin{figure*}
\includegraphics[width=\textwidth]{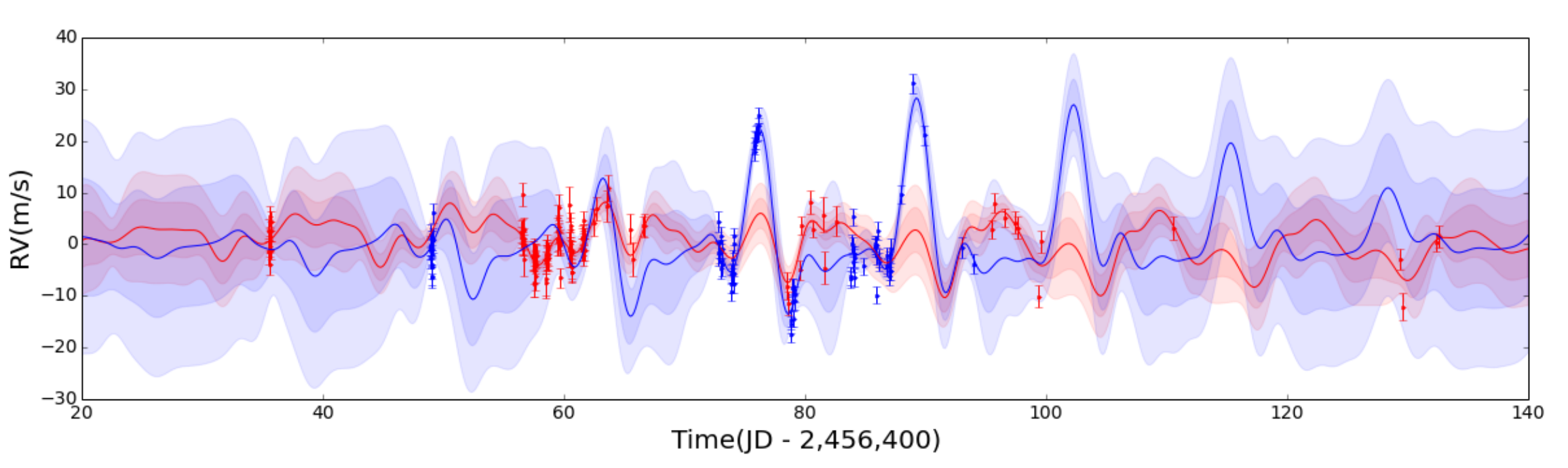}\label{fig:RVFitQP}
\caption{RV of Kepler-78 versus time, measured by Keck-HIRES and HARPS-N. The HIRES and HARPS-N data are plotted as blue and red points respectively, with errors in RV shown. A Keplerian orbit signal with the calculated best-fit Doppler amplitude has been subtracted from the data. The colored lines correspond to the Gaussian process regressions with best-fit kernel parameters of the correspondingly colored RV measurements, where the shaded regions correspond to 1-$\sigma$ and 2-$\sigma$ uncertainties. Both datasets have been fitted with a single quasi-periodic kernel operator with common period, roughness, and lengthscale ($\theta$, $w$, and $\lambda$) hyperparameters, but separate covariance amplitude and white noise ($h$ and $\sigma$) parameters.}
\label{fig:bestGPfit}
\end{figure*}

The best-fit GP kernel hyperparameters and the Keplerian Doppler amplitude are found via MCMC exploration of parameter space. We simultaneously fit the Keplerian planetary signal and a GP regression to each RV dataset with the Doppler amplitude and kernel hyperparameters as free parameters in the MCMC chain. The \texttt{emcee} package contains an Affine-invariant Monte Carlo Markov Chain Ensemble sampler, which evaluates the likelihood of the GP kernel hyperparameters to the measurement residuals after a Keplerian planetary signal has been subtracted \citep{foremanmackey2013}. This is done for a plethora of steps in free parameter space. We draw the best-fit GP kernel hyperparameters and planetary Doppler amplitude as well as their errors from the posterior distributions generated through this MCMC exploration of parameter space, with 1-$\sigma$ error corresponding to 68$\%$ confidence intervals in the MCMC posterior distributions. We plot the data and best-fit GP regression + Keplerian models in Figure \ref{fig:bestGPfit}.

After the MCMC posterior distributions have been created, we calculate a Gelman-Rubin statistic for each parameter to ensure that the parameter chains have converged. Convergence is deemed adequate when the G-R statistic $<$ 1.01 \citep{gelman1992}. In addition, we track the average and median log likelihood at each step in the chain. At first, the chains move to higher and higher likelihood parameter space. Once the best-fit solution is found, the chains begin to take more and more unlikely steps, exploring parameter space more completely, and pushing the average likelihood below the median value. The step at which the average log likelihood begins dipping below the median log likelihood indicates the transition from burn-in period to the exploration of parameter space around the maximum likelihood solution, and we remove the steps taken during burn-in from our analysis \citep[e.g.][]{knutson2008}. To check that the resultant fit is a good description of the data, we compute a $\chi^2_{\mathrm{red}}$ for the best-fit model of each kernel variant. Finding $\chi^2_{\mathrm{red}} \approx 1$ confirms that the quasi-periodic GP model and errors describe the data well.

We allow as many of the MCMC parameters as possible be minimally constrained. We find that whenever the periodic or quasi-periodic kernel is used, and the photometric $\theta$ value is not applied as a prior, the RV noise model is able to recover periodicity consistent with the stellar rotation period. This supports our assumption that the stellar surface features are the predominant signal seen in the RV measurements. Applying a Gaussian prior to this value allows us to better constrain all other MCMC parameters. Physical boundary conditions were also placed on some hyperparameters to prevent the MCMC chain from moving into unphysical parameter space (see Table \ref{tbl-qmp} for details). 

\begin{figure}
\plotone{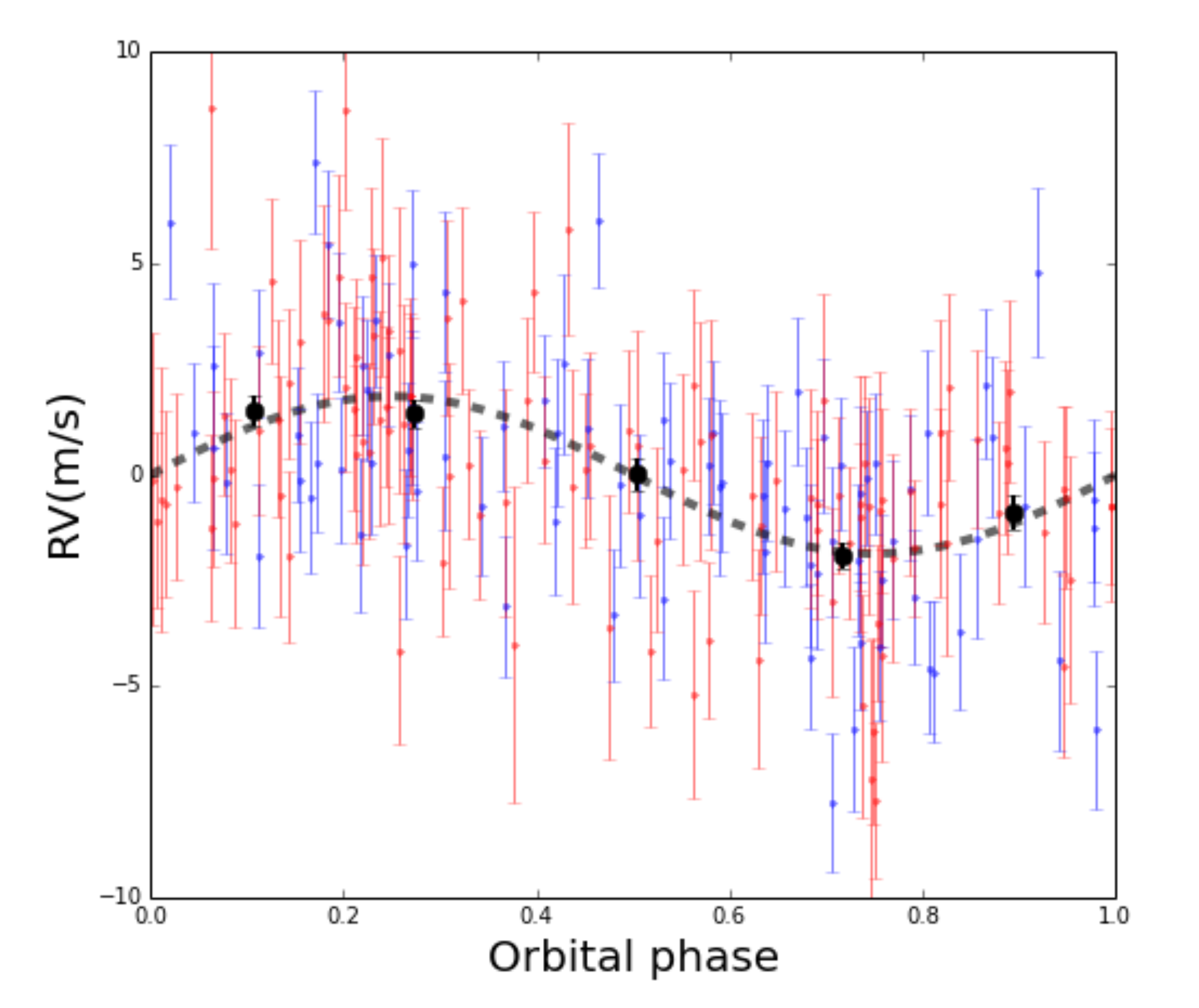}
\caption{The residuals of the HIRES and HARPS-N data after the quasi-periodic GP regression to the data with the Keplerian signal subtracted is removed, phase-folded at the known orbital period of the planet. HIRES data is shown in blue, and HARPS-N data is shown in red. The planetary signal model is shown by the dotted black line, and binned data are shown by black points.} \label{fig:GPfitphase}
\end{figure}

\section{Results}

 We find a Doppler amplitude of Kepler-78b of \Dopamp, corresponding to a mass of \mass\ using the quasi-periodic kernel with common temporal hyperparameters, excepting amplitudes and white noise terms, for the two RV datasets. We find that the quasi-periodic kernel provides the best fit to the Kepler photometry based on visual signal inspection, and the validity of the fit is confirmed through $\chi^2_{\mathrm{red}}$ calculation.  Based on a Bayesian analysis, we find the strongest evidence for a quasi-periodic kernel regression of the radial velocity data over simpler kernels as well, and that stronger evidence exists for one set of temporal hyperparameters to describe both radial velocity datasets as opposed to two, suggesting that the dominant signal in both RV datasets is indeed astrophysical. This supports the findings of \citet{haywood2014}. We provide the adopted model parameters for the best-fit quasi-periodic RV GP regression + Keplerian orbit models chosen in Table \ref{tbl-qmp}.


\subsection{GP kernel selection}
 
We find a planetary mass within 1-$\sigma$ of our adopted mass result for all models tested. We make the assumption that the predominant signal observed in the photometric and both RV datasets comes from the star, and thus will modulate at the stellar rotation period. We find that when we model the photometric data with a periodic or quasi-periodic GP kernel, we recover a period hyperparameter consistent with previous estimates of the stellar rotation period, supporting our assumption about the predominant photometric signal. Similarly, when we model the RV datasets with a quasi-periodic GP with no priors, we recover a period hyperparameter consistent with the stellar rotation period or its alias. Thus, we conclude that the quasi-periodic GP regression is both well motivated and effective for describing the predominant signal in our datasets.

We calculated BIC and AIC values for all of our RV models. We test the three kernels of interest both with shared hyperparameters as well as with fully independent hyperparameters in order to ensure that the variation we observe is consistent between datasets and thus astrophysical, rather than local, in origin. Furthermore, we test our squared exponential kernel models with and without a white noise term for each dataset. We report these BIC and AIC values in Table \ref{tbl-4}. 

We find that the quasi-periodic model we choose has AIC and BIC values with $\Delta$BIC $\geq -0.5$ and $\Delta$AIC $\geq 8$ between it and all other models tested. A $\Delta$BIC value of less than 2 suggests two models are indistinguishably likely, whereas a $\Delta$BIC greater than 10 suggests strong evidence for the model with a lower BIC value.  Interestingly, the BIC for the quasiperiodic and squared exponential kernels are almost identical, while there is significantly less evidence for the periodic model (BIC of periodic model 18 points higher than SE and quasi-periodic models). Furthermore, the lowest BIC of all was found for the squared exponential kernel with two independent GP regressions, with a $\Delta$BIC = $-0.5$ as compared to the chosen model. \textit{However, the AIC of the quasi-periodic, coupled GP model is significantly lower than that of either of the other models, with a $\Delta$AIC=8 between it and the next lowest AIC value, that for the quasi-periodic, uncoupled model, and $\Delta$AIC=14 for the coupled model with a squared exponential kernel. The relative likelihood of one model to another is given by $e^{-\Delta \mathrm{AIC}/2}$, indicating that the uncoupled quasi-periodic model is 1$\%$ as likely as the coupled model, and the coupled squared exponential kernel model is 0.09$\%$ as likely.} The independent GP, squared exponential kernel had a $\Delta$AIC=19 when compared to the chosen model. We also tested removing the white noise parameter from the squared exponential model, finding that doing so raised the BIC by over 20 points and the AIC by over 15. Thus, we determined that the white noise parameter was justified for all kernels. Strictly speaking, these information criteria require the assumption that the noise in the data is independent and identically distributed, which we know is not the case. However, since we use the same data for all BIC and AIC comparisons, they provide a valuable comparison between the different GP kernels. We report all relevant AIC and BIC values in Table \ref{tbl-4}.
  
\begin{figure*}
\includegraphics[width=\textwidth]{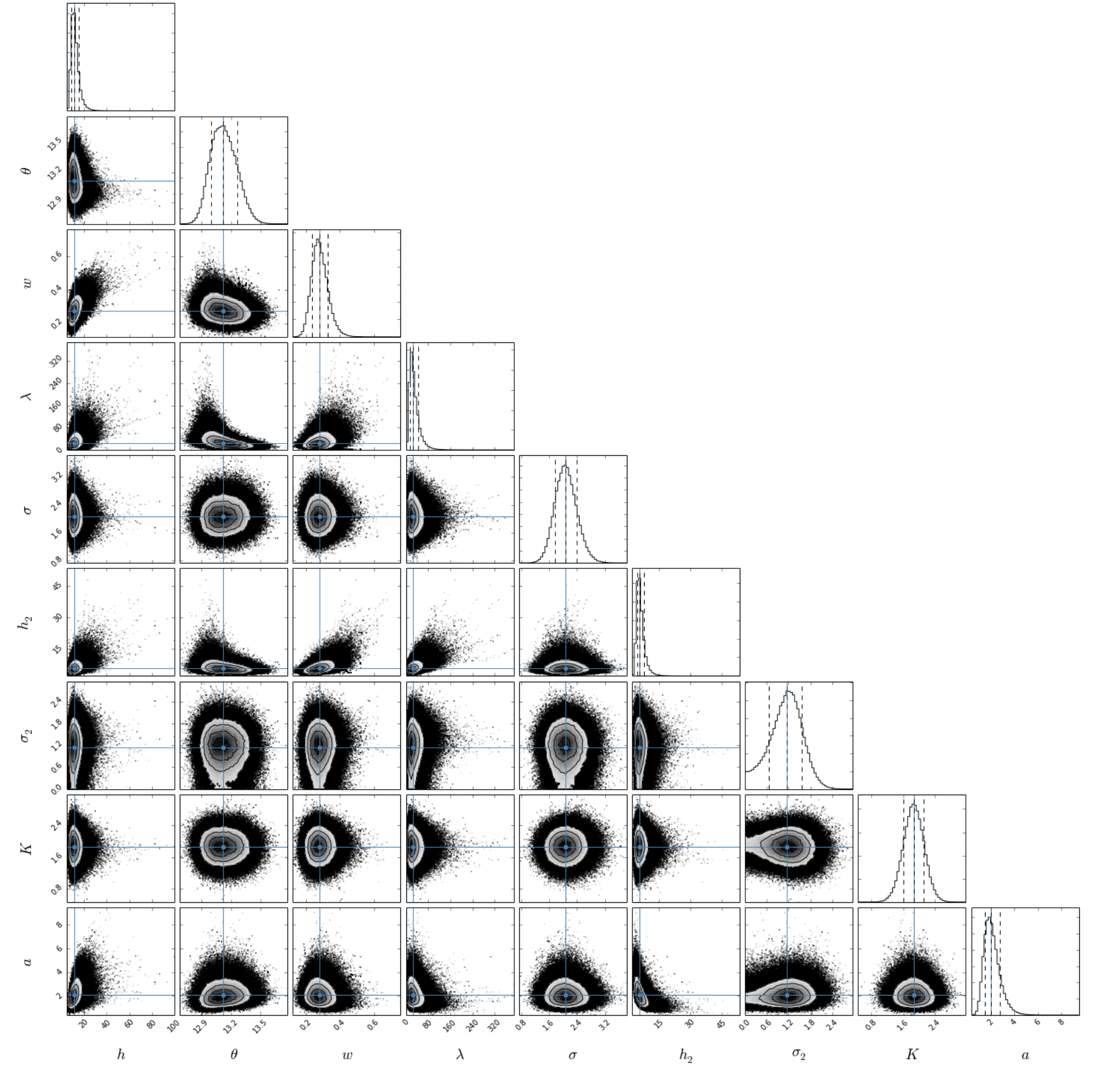}
\caption{Parameter distributions from our chosen GP + Keplerian model of the RV data plotted against each other. The median value of each parameter (blue lines) and 1-$\sigma$ ranges (dashed lines) are shown. The parameter $a = h / h_2$.}\label{fig:triangleQP}
\end{figure*}

 
 If we are modeling the stellar activity in both datasets, we would expect the hyperparameters for our simultaneous models to be the same, with the exception of the non-temporal amplitude hyperparameter $h$ and the white noise term $\sigma$, which would be different but related between the models due to the different appearance of the disk of the star in different wavelength regimes. We explored the relation of the GP kernel amplitude hyperparameters, whose relationship we evaluated by calculating $a$, the ratio of the HARPS-N amplitude to the HIRES amplitude. The convergence of $a$ to a best-fit value suggests that there is a steady relation in the ranges of the RV signals observed by HIRES and HARPS-N. Additionally, the value of $a$ is broadly consistent with the results of \citet{desort2007}, who illustrate that the expected radial velocity contribution from starspots on a K dwarf is larger at shorter wavelengths, resulting in a larger signal in HIRES than in HARPS-N, as HARPS-N has significant wavelength coverage in the 600-700nm regime that HIRES does not. We show the distribution of $a$ along with the other parameters of our MCMC calculation in Figure \ref{fig:triangleQP}. Any variation due explicitly to the correlated stellar activity should thus cause our models to have a shape (period and phase) that is consistent between the two datasets. We allow the reader to confirm this visually in Figure \ref{fig:bestGPfit}.
 
 \begin{deluxetable}{ccrrrrrrrrcrl}[h]
\tabletypesize{\scriptsize}
\tablecaption{Planetary Properties\label{tbl-2}}
\tablewidth{0pt}
\tablehead{
\colhead{Property} & \colhead{Value}
}
\startdata
Name& Kepler-78b \\
Radius, R$_{\mathrm{pl}}$ & 1.20 $\pm$ 0.09R$_{\oplus}$\\
Orbital period $P_{\mathrm{orb}}$ & 0.35500744 $\pm$ 0.00000006 days\\ 
Doppler amplitude, K & \Dopamp \\
Mass, M$_{\mathrm{pl}}$ &  \mass \\
Density, $\rho_{\mathrm{pl}}$ & \density \\
Iron fraction & \ironfrac \\
\enddata 
\tablecomments{The name, radius, and orbital period values in Table \ref{tbl-2} are taken from \cite{howard2013}. All other values have been calculated for this work.}

\end{deluxetable}


Distributions for each parameter relative to all other parameters are shown in Figure \ref{fig:triangleQP} for the quasi-periodic GP + Keplerian models. We see no clear correlation between Doppler amplitude and any other parameter tested, indicating that the Doppler amplitude observed is a real signal and not a systematic error introduced during our parameter fitting.

We place a Jeffreys prior on the period hyperparameter $\theta$ to weight shorter periods more heavily when fitting the quasi-periodic and periodic GP regression kernel hyperparameters to the photometric data \citep{haywood2014}. When we do this, we are able to recover the rotation period of the star as the period hyperparameter. We then place a Gaussian prior on this hyperparameter in the RV GP + Keplerian model constraining it to the value and errors measured in the photometric data. We also note that although the coherence scale parameter $w$ measured in the photometric data is equal to that measured in the RV data within errors, we do not place a similar Gaussian prior on this parameter, because we observe that it is correlated with both the amplitude and characteristic timescale hyperparameters, which were not necessarily related between the photometric and RV GP regression hyperparameter fits. This correlation between $h$, $h_2$, $w$ and $\lambda$ visible in Figure \ref{fig:triangleQP} is hard to interpret physically. We speculate that the positive correlation between $w$ and the $h$ parameters may arise from a connection to starspot size: as starspots grow larger, $h$ grows, any white noise present will become relatively less important, and thus the function will become smoother overall, increasing the $w$ parameter. Similarly, the characteristic timescale $\lambda$ may also increase at large amplitudes for the same reason--as the uncorrelated noise at small timescales becomes less important, the lengthscale might be weighted more heavily toward larger values. Larger spots also persist longer on the stellar surface, as spots decay on timescales proportional to their size \citep{bumba1963}. However, the fact that $w$ was consistent for the photometric and RV datasets whereas $\lambda$ was smaller for the photometric datasets despite their large differences in amplitudes may suggest otherwise. In addition, the amplitude parameters $h$ and $h_2$ are not directly related to the range of the RV shift observed, and thus difficult to interpret physically. Further tests of these parameters are necessary in order to fully characterize their relationships, such as placing new priors on the coherence scale and characteristic timescale parameters to ensure that they are not biased by noise.


\begin{deluxetable}{ccc}
\tabletypesize{\scriptsize}
\tablecaption{Adopted Quasi-periodic RV Model Parameters\label{tbl-qmp}}
\tablewidth{0pt}
\tablehead{
\colhead{Name} & \colhead{Prior} & \colhead{Value}
}
\startdata
HIRES Amplitude $h$& $h > 0$ & 11.6$^{+3.7}_{-2.5}$ m s$^{-1}$ \\
HARPS-N Amplitude $h_2$& $h_2 > 0$ & 5.6$^{+2.0}_{-1.3}$ m s$^{-1}$\\
Amplitude ratio $a$ & - & 2.0$^{+0.8}_{-0.5}$ \\
Period $\theta$ & $e^{-\frac{(13.26-\theta)^2}{2(0.12)^2}}$ & 13.12 $^{+0.14}_{-0.12}$ days\\
Coherence scale $w$ & - & 0.28$^{+0.05}_{-0.04}$\\ 
Characteristic timescale $\lambda$ &  $\lambda > 0$ & 26.1 $^{+19.8}_{-11.0}$ days\\
HIRES white noise $\sigma_{\mathrm{jitter}} $ & $\sigma > 0$ & 2.1$^{+0.3}_{-0.3}$ m s$^{-1}$ \\
HARPS-N white noise $\sigma_{\mathrm{jitter}} $ & $\sigma > 0$ & 1.1$^{+0.4}_{-0.5}$ m s$^{-1}$\\
Doppler amplitude, K & - & \Dopamp \\
\enddata 
\end{deluxetable}



\begin{deluxetable*}{llllllllrcccccc}
\centering
\tabletypesize{\scriptsize}
\tablecaption{Model MCMC Fit Diagnostics and Results\label{tbl-4}}
\tablewidth{\textwidth}
\tablehead{
\colhead{Kernel function} & \colhead{K (m s$^{-1}$)}  & \colhead{BIC value (2GPs)\tablenotemark{b}} & \colhead{AIC value (2GPs)\tablenotemark{b}} &\colhead{Comment}}
\startdata
Quasi-periodic (adopted)  \tablenotemark{a}& \Dopamp & 1065.3 (1079.6) & 1039.2 (1047.0) & with white noise, common $w$, $\theta$, $\lambda$ \\
& & & & but different $h$, $\sigma$ parameters\\
\\
Periodic \tablenotemark{a} & 1.82 $\pm$ 0.29 m s$^{-1}$ & 1083.8 (1090.6) & 1067.5 (1061.2) &  with white noise, common $w$, $\theta$, \\
& & & & but different $h$, $\sigma$ parameters  \\
\\
Squared exponential &1.92 $\pm$ 0.27 m s$^{-1}$  & 1066.4 (1064.8) & 1053.4 (1058.4) & with white noise, common $\lambda$ but \\
 & & & &  but different $h$ parameters\\
\enddata 

\tablenotetext{a}{For the periodic and quasi-periodic kernel models, the period hyperparameter $\theta$ is constrained by a Gaussian prior. This prior is shaped by the posterior distribution of the corresponding photometric GP hyperparameter. When this prior was not applied, the best-fit period found in the RV data was consistent with the stellar rotation period, but Doppler amplitude errors were larger.}
\tablenotetext{b}{We measured the AIC and BIC values for all models with common RV GP hyperparameters and with 2 independent hyperparameter sets, which we call 2GPs here. We also measured BIC and AIC of the 2GP squared exponential kernel without a white noise term but do not include them in this table.}

\end{deluxetable*}

\section{Interpretation and Discussion}

With our measured mass of \mass\ and a radius of 1.20 $\pm$ 0.09 R$_{\oplus}$ (H13), Kepler-78b has a bulk density of \density, suggesting a rocky composition similar to Earth ($\rho_\oplus = 5.52$ g cm$^{-3}$).  Using the two-component, rock-iron models from \citet{fortney2007}, we estimate an iron fraction of \ironfrac, consistent with an Earth-like composition \citep[iron mass fraction of 0.319,][]{mcdonough1995}. These simplified models consider Kepler-78b as an iron core surrounded by a rocky mantel, and account for compression that is important for higher planet masses.  We ignore the effect of an atmosphere on the radius of Kepler-78b due to its equilibrium temperature of 1500-3000 K \citep{sanchis-ojeda2013}.  

\begin{figure}
\vspace{0.2cm}
\plotone{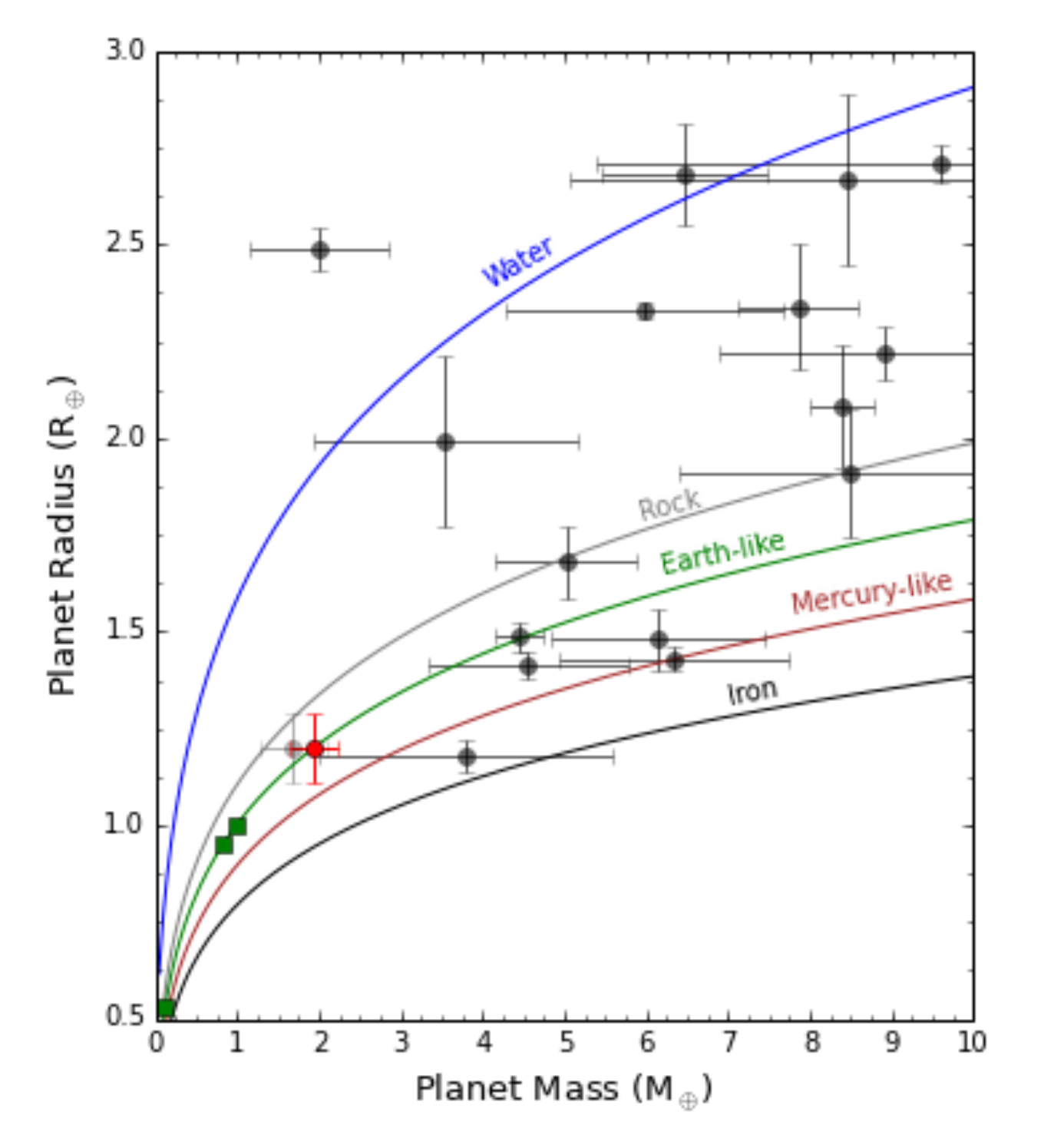}
\caption{A graph of mass versus radius for rocky planets. The constraints on Kepler-78b are shown by the red point. The estimation of Kepler-78b by H13 is shown as the light gray error bars. Composition curves ranging from a pure water (blue) to pure iron composition (black) have been plotted. Earth-like (67$\%$ rock, 33$\%$ iron) and Mercury-like (67$\%$ iron, 33$\%$ rock) compositions are denoted by green and brown curves, respectively. Solar system planets are shown as green squares. Other well-characterized exoplanets are plotted as black points. Exoplanet masses, radii, and their associated errors are from the Exoplanet Orbit Database (http://exoplanets.org; downloaded on 23 August 2014). Planets with fractional mass uncertainties of over 50$\%$ are not shown.} \label{fig:test}
\end{figure}

We now speculate about the implications of these measurements to the composition of Kepler-78b. In Figure \ref{fig:test}, we illustrate that the best-fit composition of Kepler-78b is indistinguishable from the composition of Earth. However, Kepler-78b's range in iron mass fraction stretches from almost purely rock (a Moon-like composition) to 60$\%$ iron (a Mercury-like composition). This is consistent with a range of rocky solar system planets but distinguishable from other solid celestial bodies (such as M-type asteroids, which are almost 100$\%$ iron by mass, or comets, made up of rock and ice with no iron at all). Since its best-fit composition and mass are more similar to Earth than any other solar system body, Kepler-78b might have had a formation process that was similar to Earth's. The proximity of Kepler-78 to its host star (0.009 AU) makes it unlikely that it formed in its current location, and migration to its current orbit is likely.


The improvement in mass determination of this model relative to H13 is shown in Figure \ref{fig:test}. Despite the fact that the mass measured by this method is a 6.5-$\sigma$ measurement compared to H13's 4-$\sigma$ and P13's 6-$\sigma$ detection, the new measurement of Kepler-78b's density is as precise as those obtained by either of the two competing teams in 2013, who estimated its density to be 5.3$^{+2.0}_{-1.6}$ (H13) and 5.57$^{+3.0}_{-1.3}$ g cm$^{-3}$ (P13). Relative to P13, the errors on the density in this study are 15$\%$ smaller, although only marginally smaller than that of H13. This is because even though the new mass measurement is somewhat more precise, the error on the density is driven three times more strongly by the error on the radius than the mass. Higher cadence photometry of the planetary transit would allow for ingress and egress of the system to be observed, breaking the degeneracy between the impact parameter and transit depth. Estimates of the planet's density could also be improved with more accurate measurements of the stellar mass, as converting RV measurements into a planetary mass is directly dependent on the stellar mass.  


\citet{hatzes2014} tested several traditional models on the same RV data analyzed here to check the robustness of the planetary detection and verify whether the planet could be found without previous transit knowledge. A range of Doppler amplitudes of the planet from 1.31 to 1.96 m s$^{-1}$ are reported, which is broadly consistent with the result of this work as well as that of H13 and P13. \citet{hatzes2014} reports the planet can be identified without prior knowledge of the system using a modified version of the parametric method originally tested by P13, the floating chunk offset model, to analyze both datasets, and reports a mass of 1.31 $\pm$ 0.24 M$_{\oplus}$, inconsistent at the 1-$\sigma$ level with the mass calculated in both this study and the previous studies. The spread in the Doppler amplitudes reported by \citet{hatzes2014} illustrates that slight differences in the choice of noise model has a significant effect on the planetary signal extracted, and that the error on the planetary mass measurement has likely been underestimated. Thus, by starting with the simplest possible nonparametric descriptions of the data, we find a description of the data that is minimally complex yet strongly supported by Bayesian information criteria. Furthermore, the structure of the best fit GP kernel as well as the success of sharing kernel parameters to describe both datasets supports our belief that we have robustly measured the astrophysical variability of the Kepler-78 system, and can obtain a mass measurement for Kepler-78b that is consistent with previous work.

\section{Summary}

We performed a combined analysis of the photometric and RV data of Kepler-78 in order to better extract the RV signal of the planet Kepler-78b. We fit the data using Gaussian process regression, and test three simple kernel configurations. After testing multiple models, we find the strongest evidence for squared exponential kernel models, and a quasi-periodic kernel model in which the hyperparameters for the GP regression for each RV dataset are coupled (aside from the amplitude hyperparameter and the white noise term). We prefer the quasi-periodic model due to evidence indicated by AIC calculation and because it supports our understanding of the physical origins of the observed signals. We measure the Doppler amplitude and therefore the mass of Kepler-78b to 6.5-$\sigma$ significance, comparable to or better than all previous mass measurements of this planet. We constrain the iron mass fraction of the planet to \ironfrac, illustrating that Kepler-78b is most likely Earth-like in composition. 

The analysis done in this work (and previous studies of this system) is possible because the orbital period of the planet P$_{\mathrm{orb}}$ is an order of magnitude smaller than and not a harmonic of the rotation period P$_\mathrm{rot}$ of Kepler-78. This made separation of the signals related to the stellar rotation and the signal related to the planetary orbital period possible. If the planetary period was a larger fraction or a harmonic of the stellar rotation period, deconvolving the signal due to the star and the signal due to the planet would be much more difficult.

\textit{The true benefit of this analysis technique comes from its nonparametric nature.} This analysis is particularly powerful in that even if the actual nature of the noise being modeled is unclear, a GP model can still be used to explore the nature of the noise and identify the most evident components. Due to the success of extracting a quasi-periodic signal at the rotation period of the star in independent datasets, we conclude that the predominant noise signal in the RV datasets likely comes from the stellar activity of Kepler-78. The variability of this noise, due to the growth and decay of active regions on the stellar surface, cannot be fully parametrized with the information we have. Thus, testing the nonparametric Gaussian process noise model is a valuable exploration after earlier parametric and nonparametric models of the RV activity. Furthermore, the use of the \textit{Kepler} photometry as a prior on our estimate of the RV activity makes our Gaussian process analysis especially valuable when photometric data is more readily available for a system than spectroscopic data, as is often the case. 


We plan to further test the robustness of this technique by analyzing other RV datasets of exoplanetary systems. This method is particularly useful for analyzing stellar RV datasets over long time baselines which cannot be modeled easily because of the spot evolution on a timescale less than an order of magnitude larger than the stellar rotation period (as determined by the autocorrelation function analysis done by P13). Such scenarios can be easily described with a quasi-periodic GP. In addition, since kernel functions can be combined, any sort of physical combination of periodic and linear or exponential signals can be modeled with the GP, indicating that it could be particularly useful in describing other noise modes seen in radial velocity studies as well as the already well-understood stellar surface signals. We hope to use this technique on a system with contemporaneous photometry and spectroscopy to explore the relationship between photometric and spectroscopic signals of exoplanetary systems.

\acknowledgments{We thank the anonymous referee for their suggestions. We also thank Suzanne Aigrain, B.\,J.\ Fulton, Conor McPartland, and Maxwell Service for helpful discussions.  
A.\,W.\,H.\ acknowledges NASA grant NNX12AJ23G.  
We gratefully acknowledge the efforts and dedication of the Keck Observatory staff and extend special thanks to those of Hawai`ian ancestry on whose sacred mountain of Mauna Kea we are privileged to be guests.  Without their generous hospitality, the Keck observations presented herein
would not have been possible.}

\bibliographystyle{apj}
\bibliography{Kepler78b-Grunblatt-rev1}

\end{document}